\documentclass[usenatbib,usegraphicx]{mn2e}

\newcommand{\gyr}{\mbox{$\rm\,Gyr$}}
\newcommand{\myr}{\mbox{$\rm\,Myr$}}

\newcommand{\msun}{\mbox{$\,M_\odot$}}
\newcommand{\kms}{\mbox{${\rm\,km\,s}^{-1}$}}
\newcommand{\beq}{\begin{equation}}
\newcommand{\eeq}{\end{equation}}
\newcommand{\arcdeg}{\mbox{$^\circ$}}
\newcommand{\gtrsim}{\goa}
\newcommand{\lesssim}{\loa}
\newcommand{\apj}{ApJ}
\newcommand{\apjl}{ApJ}
\newcommand{\apjs}{ApJS}
\newcommand{\aj}{AJ}
\newcommand{\mnras}{MNRAS}
\newcommand{\aaps}{A\&AS}
\newcommand{\aap}{A\&A}
\newcommand{\pasp}{PASP}

\title[Triaxiality in NGC 4365]
{Long-Lived Triaxiality in the Dynamically Old Elliptical Galaxy
NGC 4365: A Limit on Chaos and Black Hole Mass}

\author[Statler et al.]
{Thomas S. Statler$^1$\thanks{E-mail: statler@ohio.edu},
Eric Emsellem$^2$,
Reynier F.\ Peletier$^{2,3}\thanks{Present address: Kapteyn Institute,
PO Box 800, 9700 AV Groningen, The Netherlands}$
and Roland Bacon$^2$\\
$^1$Department of Physics and Astronomy, 251B Clippinger Research
Laboratories, Ohio University, Athens, OH 45701, USA\\
$^2$Centre de Recherche Astronomique de Lyon, Observatoire de Lyon,
9 Avenue Charles-Andr\'e, 69230 Saint-Genis-Laval, France\\
$^3$Department of Physics and Astronomy, University of Nottingham,
University Park, Nottingham NG7 2RD, UK}

\begin{document}

\date{Accepted 2004 May 25; in press.}

\pagerange{\pageref{firstpage}--\pageref{lastpage}} \pubyear{2004}

\maketitle

\begin{abstract}

\label{firstpage}

Supermassive black holes in the centres of giant elliptical galaxies
are thought to be capable of inducing chaos and eliminating or preventing
triaxiality in their hosts if they are sufficiently massive. We address
whether this process operates in real systems, by modeling the stellar
kinematics of the old elliptical NGC~4365. This galaxy has a mean stellar
population age $>12\gyr$ and is known for its kinematically decoupled
core and skew rotation at larger radii. We fit the two-dimensional mean
velocity field obtained by the SAURON integral-field spectrograph, and
the isophotal ellipticity and position-angle profiles, using the velocity
field (VF) fitting approach. The models constrain the system's intrinsic
shape between $0.03$ and $0.5$ effective radii, as well as its orientation
in space. We find NGC~4365 to be strongly triaxial ($\langle T \rangle
\approx 0.45$) and somewhat flatter than it appears ($\langle c/a \rangle
\approx 0.6$). Axisymmetry or near axisymmetry ($T<0.1$) is ruled out at
$>95\%$ confidence for $1\farcs 6 < R < 3\farcs 2$ ($0.03 < R/r_e < 0.06$),
and at $>99\%$ confidence
at larger radii. There is an indication of an outward triaxiality gradient.
The line of sight is constrained to two narrow bands on the viewing
hemisphere. In the most probable orientation the long axis points roughly
toward the observer, extending to the southwest in projection. The stellar
population age implies that strong triaxiality has persisted for hundreds
of dynamical times. This rules out black holes $>3 \times 10^9 \msun$, which
numerical simulations indicate would either have globally axisymmetrized
the galaxy or made the inner several arcseconds spherical. The
$M_{\rm BH}$-$\sigma$ relation predicts $M_{\rm BH} \approx 4\times 10^8\msun$,
which would probably not preclude long-lived triaxiality and is consistent
with the observations. There must also be an unequal population of direct
and retrograde long-axis tube orbits outside the kinematically decoupled
core. This, combined with the small isophotal twist, limits the rate of
figure rotation (tumbling) about the short axis, and places corotation at
$>8$ effective radii. NGC~4365 lends support to a picture in which
supermassive black holes, though omnipresent in luminous giant elliptical
galaxies, are not massive enough to alter their global structure
through chaos.

\end{abstract}


\section{Introduction\label{S.INTRO}}

There is growing recognition that the dynamical structure of hot stellar
systems---elliptical galaxies and the bulges of spirals---is
intimately connected with the supermassive black holes that reside at their
centres. The most compelling evidence for this connection is the
tight correlation between black hole masses and the velocity dispersions of
their host galaxies measured far from the black hole. The
$M_{\rm BH}$-$\sigma$ relation \citep{FM,NukerMsigma} indicates that central
black holes somehow ``know'' about the properties of their hosts, and
has provoked speculation that host dynamics control the growth of the central
object \citep{ElZant,Adams}. But the exact mechanisms by which this
control might be exerted are still obscure.

By contrast, a mechanism by which a central black hole may, in a turnabout
move, affect the structure of its host {\em is\/} understood, or at
least identified. Important families of regular orbits that pass close
to the centre of the system can be rendered chaotic by scattering off the
central mass \citep{LakeNorman, Gerhard85, Gerhard86}.
It has thus been suggested that growth of
a black hole can lead to the destruction of bars in spirals
\citep{Hasan,Norman} or to the elimination of triaxiality in ellipticals
\citep{Valluri,MQ}. In this case, however, what is lacking is direct
evidence that this mechanism has played a role in real systems.

When central point masses or density cusps are added to triaxial potentials,
box orbits are transformed to stochastic orbits or resonant ``boxlets''
\citep{Schwarzschild}.\footnote{For some special counter-examples see
\citet{Sridhar97,Sridhar00}.}\ Despite some speculation
that the loss of boxes would preclude nearly all triaxial equilibria
\citep{MerrittScience}, it is now
understood that stochastic orbits can contribute constructively to
maintaining the triaxial figure \citep{Schwarzschild, MF}.
The important questions now
focus on the chaotic mixing time scale \citep{Siopis}, and whether, for
practical purposes, one can consider there to be many distinct, slowly
diffusing, stochastic orbits at each energy, rather than only one such
orbit occupying the entire Arnol'd web \citep{MF}. Studies of individual
orbits \citep{Valluri,Poon1} as well as evolutionary $N$-body simulations
\citep{MQ,Sellwood} agree that the box-orbit region of phase space
becomes fully chaotic when the central point mass reaches $\sim 1\%$ of the
total system mass. This transition happens somewhat earlier for more elongated
figures or steeper cusps. At lower masses, a fully chaotic zone appears
over a limited range of energies, affecting a region whose enclosed stellar
mass is of order 10 times that of the central object \citep{Poon2,HB}.

The current $M_{\rm BH}$-$\sigma$ relation \citep{TremaineMsigma}, coupled
with the Fundamental Plane \citep{Djorg,Dressler},
implies that black holes will typically fall about a factor of 10 short of
the 1\% mass needed to induce global chaos and a major restructuring of
the host galaxy. If this is universally the case, triaxiality should persist
over long times, and we should expect to find many old, triaxial ellipticals.
However, \citet{Valluri} argue that the threshold for
chaos should be proportionally lower in lower luminosity galaxies, which have
steeper cusps and shorter dynamical times, and suggest that galaxies
fainter than around $M_B \approx -20$ should have been axisymmetrized
by chaos. Consequently, determining the actual abundance of triaxial
systems as a function of luminosity will be an important check on our 
understanding of the masses of black holes and their influence on their hosts.

Determining the true shapes of elliptical galaxies
cannot be done by photometry alone. Kinematic data are essential
\citep{Binney85,FIZ}, and the ``traditional'' major and minor axis kinematic
profiles are not enough. \citet{Statler94a} advocated a minimum of 4
long-slit position angles for a credible constraint on triaxiality, a
requirement that is prohibitively costly in telescope time. Happily,
this has become a moot point with the advent of integral-field
spectrographs. The SAURON instrument \citep{BaconSauron} was developed
specifically to map the kinematic and spectral properties of galaxies,
with full two-dimensional coverage on the sky. SAURON has been used to
observe a sample of 72 E, S0, and Sa galaxies \citep{deZeeuwSauron},
producing an extensive catalog of kinematic and line-index measurements.
In this paper we fit triaxial models to the SAURON kinematic
dataset for the old elliptical NGC~4365, in order to estimate its
intrinsic shape and constrain its orientation relative to the line of sight.

The properties of NGC~4365 have recently been summarized by
\citet{DaviesSauron}. It is a luminous giant elliptical ($M_B=-21.05$ for
a distance modulus of $31.55$) with a very shallow central cusp. The
negative logarithmic slope of the inner surface brightness profile is $\gamma
\approx 0.13$ \citep{Rest}. Photometrically, the galaxy has
near-constant ellipticity $\epsilon \approx 0.26$, and a modest
isophotal twist of about $10\arcdeg$ from $0.02$ to 2 effective radii
($r_e=57\arcsec$). The isophotes are discy in the inner $4\arcsec$ and
boxy farther out. NGC~4365 is probably best known for its kinematics, having
both a kinematically decoupled core and skew rotation at larger radii.
The apparent rotation is about the photometric minor axis in the core,
and $\sim 40\arcdeg$ from the major axis at $20\arcsec$ from the
centre. These kinematic signatures were recognized by \citet{Surma} as
hallmarks of triaxiality. From the SAURON line index maps, \citet{DaviesSauron}
determine that the stellar population in NGC~4365 has a mean age of $14\gyr$,
both in and outside the kinematically distinct core,
with at most a few percent admixture of an intermediate-age population in
the innermost $1\farcs 6$. They find that putting 6\% of the mass in a
5-Gyr-old population with the same metallicity as the rest of the galaxy
can account for the slightly elevated central H$\beta$ line-strength.
\citet{DaviesSauron} conclude that the galaxy has been largely
unchanged in the last $12\gyr$. However, there is also evidence of a
population of intermediate-age (2--8 Gyr old) globular clusters
\citep{Puzia,Larsen}, suggesting that a minor merger event a few billion
years in the past may have modified the structure of the galaxy and
possibly contributed to the kinematically decoupled core.

\begin{figure*}
\includegraphics[width=5in]{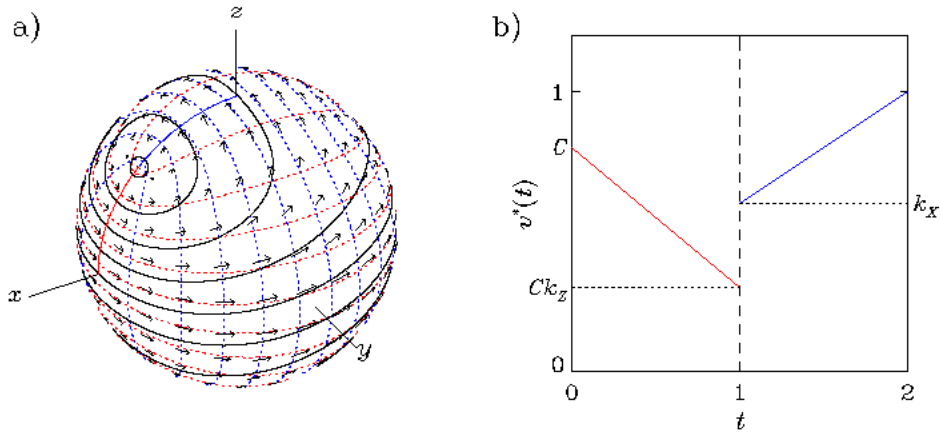}
\caption{Construction of a velocity field on one fiducial shell. ($a$) A
confocal coordinate system linked to the model triaxiality (in this example
$T=0.5$) represents the streamlines of short-axis tubes ({\em red dotted
lines\/}) and long-axis tubes ({\em blue dotted lines\/}). Given a
density distribution, specifying the velocity crossing an arc in the
$(x,z)$ plane ({\em heavy colored line\/}) then determines the mean flow
field everywhere ({\em black vectors and streamlines\/}). The boundary
arc is divided into two segments ({\it red, blue\/}) describing the
contribution of short-axis and long-axis tubes, respectively. ($b$) The
function $v^\ast(t)$, which determines the velocity along the boundary arc,
showing the definition of the constants $C$, $k_X$, and $k_Z$. The example
in ($a$) has $C=1$, $k_X=k_Z=0$.
\label{f.streamlines}}
\end{figure*}

In \S\ \ref{S.METHOD} below we describe the methods we use to constrain the
shape and orientation of the galaxy, including the nature of the models,
the handling of the observational data, and the statistical formalism. \S\
\ref{S.RESULTS} presents the basic results. We find that NGC~4365 is highly
triaxial, with axisymmetry ruled out at better than 99\% confidence between
approximately $2\arcsec$ and $30\arcsec$. We find the line of sight to be
constrained to two fairly narrow strips on the hemisphere of viewing angles.
We are also able to constrain, to a limited degree, the contributions of
the main tube orbit families to the mean velocity field.
In \S\ \ref{S.DISCUSSION} we show that the persistent triaxiality of this
old system rules out black holes larger than a few $10^9\msun$, but is
probably consistent with predictions of the $M_{\rm BH}$-$\sigma$ relation.
We also use the predominance of long-axis tubes in the outer part of the
system and the lack of a strong photometric twist to limit the rate of
figure rotation. We briefly discuss the implications
for orbit-based (Schwarzschild) and moment-based (Jeans) models, before
summarizing the principal
results in \S\ \ref{S.SUMMARY}.

\section{Method\label{S.METHOD}}

We determine the intrinsic shape of the galaxy by fitting 
models to the mean line-of-sight velocities and the radial profiles of
isophotal ellipticity and major-axis position angle. We do not make use
of the dispersion or the higher moments (except for a $v/\sigma$
constraint described below), because our goal is not to determine the
full three-dimensional mass profile. That is a job better left to the more
sophisticated---but far slower---Schwarzschild method. Our goal is only to
constrain the shapes of the isodensity surfaces and the orientation of
the galaxy relative to the line of sight. We can extract this information
from the mean rotation and photometry because the mean velocities in a
non-tumbling triaxial galaxy arise from the populations of stars on
short-axis (Z) and long-axis (X) tube orbits, and the geometry of these orbits
is determined almost entirely by the triaxiality of the potential.

\subsection{Models\label{s.models}}

The velocity field (VF) fitting approach has been described in
previous papers \citep[and references therein]{Statler3379,SDS}.
The models are based on analytic solutions to the equation of continuity.
Though the models are approximate, their calculation is extremely fast,
making it possible to explore a far larger parameter space than can be
covered by more computationally intensive orbit-based methods. We sketch
the essentials here, and explain certain details in the Appendix.

\begin{figure*}
\includegraphics[width=6in]{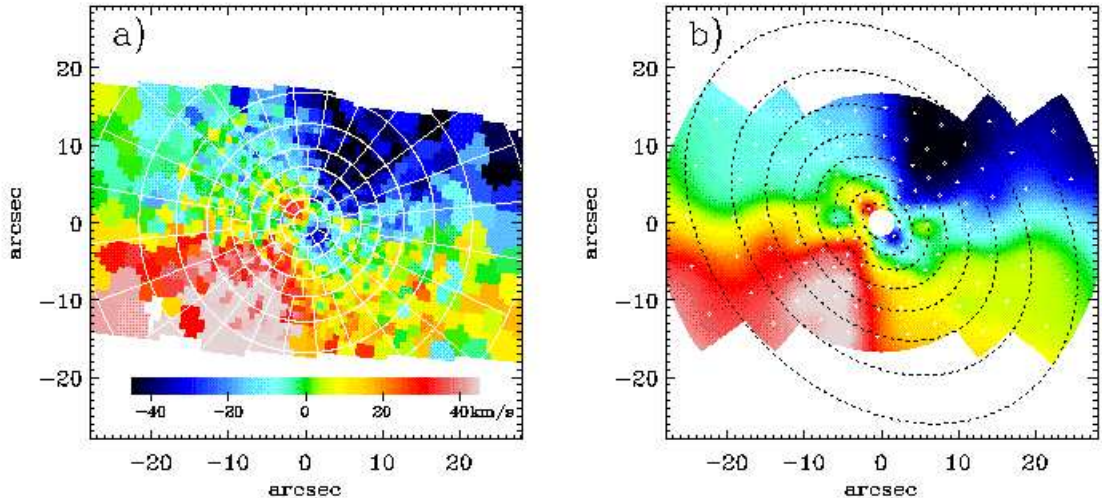}
\caption{($a$) SAURON mean velocity field for NGC~4365. Velocities are
computed from spectra averaged over Voronoi cells, and are corrected for
the $h_3$ Gauss-Hermite term inside $R=7\farcs 2$. The color bar indicates
the velocity scale relative to systemic.
Overplotted ``spider web'' shows the polar grid on to which the data are
rebinned for model fitting. ($b$) Rebinned velocity field, antisymmetrized
by folding and averaging points on opposite sides of the centre. {\em White
points\/} mark the bin centres. Colors show a continuous velocity
field interpolated from the rebinned data.
The whited-out central region is excluded from the fit.
{\em Dotted lines\/} show $V$-band isophotes. This and all similar figures
have North up and East to the left.
\label{f.data}}
\end{figure*}

The figure of the galaxy is assumed to be stationary, and the principal
axes of the isodensity surfaces are assumed to be aligned throughout.
For each projected radius at which there are data to be fitted, we
calculate an internal (3-dimensional) mean VF by solving the
equation of continuity on a deprojected shell. This solution involves
the assumed density figure of the model as well as the geometry of the
streamlines of the mean flow. To approximate the latter, we assume that
the mean motions ({\it not\/} the single-particle orbital velocities) of
X and Z tubes follow coordinate lines in a confocal coordinate system that
is keyed to the triaxiality $T$ of the mass distribution\footnote{The
triaxiality parameter $T\equiv
(a^2-b^2)/(a^2-c^2)$, where $a$, $b$, and $c$ are the long, middle, and short
axes.}\ at the radius of the shell. We have verified the accuracy of this
approximation for a wide variety of tube orbits numerically integrated
in logarithmic potentials of constant $T$ \citep{Anderson}. For
potentials with large triaxiality gradients, linking the streamlines to
the local $T$ is probably inaccurate, but this is moot here
since we find no evidence for a large gradient in NGC~4365.

Figure~\ref{f.streamlines}a shows the geometry of the streamlines and the
internal velocity field, drawn on a spherical shell in one model.
Triaxial systems can also have nonzero radial mean motions, due to the
elongation of tube orbits contrary to the elongation of the potential.
We compute each velocity field both on a spherical shell and on an
ellipsoidal shell intended to overestimate the expected radial motions, and
average the results [see \S\ 2.4 of \citet{Statler94a} for details].

The solution of the continuity equation also requires a boundary condition,
which is specified at each radius as the mean velocity along an arc in the
$(x,z)$ plane (Fig.\ \ref{f.streamlines}a). This boundary splits into two
segments, the upper determining the contribution from X tubes, and
the lower the contribution from Z tubes. Specifying the
velocity on the boundary can be seen as equivalent to specifying the
orbit populations in the model---or more precisely, that part of the
orbit population that determines the observable velocity field.
We write the absolute magnitude of the boundary velocity as $v^\ast(t)$,
where $t$ is a scaled angular variable. On a spherical shell, $t$ is
related to the polar angle $\theta$ by
\beq\label{e.vstar}
t = \left\{ \begin{array}{ll}
        2 - {\sin^2 \theta \over T}, & \theta < \sin^{-1}\sqrt{T}, \\
        {\cos^2 \theta \over 1-T}, & \theta > \sin^{-1}\sqrt{T}.
\end{array} \right.
\eeq
With this definition, $0 \leq t < 1$ is the Z tube segment of the
boundary and $1 < t \leq 2$ is the X tube segment. We take
$v^\ast(t)$ to be composed of two piecewise-linear sections, described
by a Z-tube/X-tube contrast factor $C$ and two constants, $k_X$ and
$k_Z$, that describe the variation away from the symmetry
planes (Figure~\ref{f.streamlines}b). The values of $C$, $k_X$, and $k_Z$ used
in the model grid are given in the Appendix.

Projection of the model at each radius is done using a quasi-local
approximation. Isophotes are assumed to have the shape of the projected
iso-luminosity-density surfaces at the mean deprojected radius; this
assumption is accurate in the absence of large shape gradients. The
velocity field is projected assuming that the three-dimensional flow
is similar at radii that contribute significantly to the projection integral
(i.e., close to the tangent point), scaling with radius as $r^{-\ell}$, 
and that the luminosity density at these radii scales as $r^{-k}$. In
practice we have found in all previous applications that the results are
very insensitive to the values of $k$ and $\ell$, and so in this paper we
have simply adopted $k=3$ and $\ell=0$.

Finally, we assume that the iso-mass-density and iso-luminosity-density
surfaces have the same shape at each radius. This is not as
strong as assuming that mass follows light; our assumption is only
that $M/L$ can be expressed as a function of local density. But this is
an unimportant distinction for this application to NGC~4365, since all of
the data come from inside the effective radius where the dark matter
contribution is probably negligible.

\subsection{Data\label{s.data}}

NGC~4365 was observed with SAURON on the 4.2 m William Herschel
Telescope in 2000 March. The observations and initial data reduction
are described by \citet{DaviesSauron} and \citet{BaconSauron}. More recently,
the data have been re-reduced in a homogeneous way for the entire
SAURON sample \citep{EmsellemSauron}. This re-reduction includes a new
spatial binning based on an adaptive Voronoi tessellation guaranteeing
a minimum signal-to-noise ratio of 60 per bin \citep{Cap03}. The data set
used in this paper is a penultimate version which has a slightly
lower signal-to-noise ratio per bin, but which is virtually indistinguishable
from the final version.

From the re-reduced data set we take the $V$, $\sigma$, and $h_3$ parameters
describing the Gauss-Hermite fit to the line-of-sight velocity
distribution (LOSVD) as derived by the Fourier Correlation Quotient (FCQ)
method, along with their errors. For radii $R<7\farcs 2$, where $h_3$
is significantly nonzero, we correct the mean velocity for the skewness
of the LOSVD \citep{vdMF,SSC}. These corrections are significant, and reduce
the core rotation by nearly 50\%. The corrected mean VF is
shown in Figure~\ref{f.data}a, at the full resolution of the SAURON instrument.

\begin{figure*}
\includegraphics[width=6in]{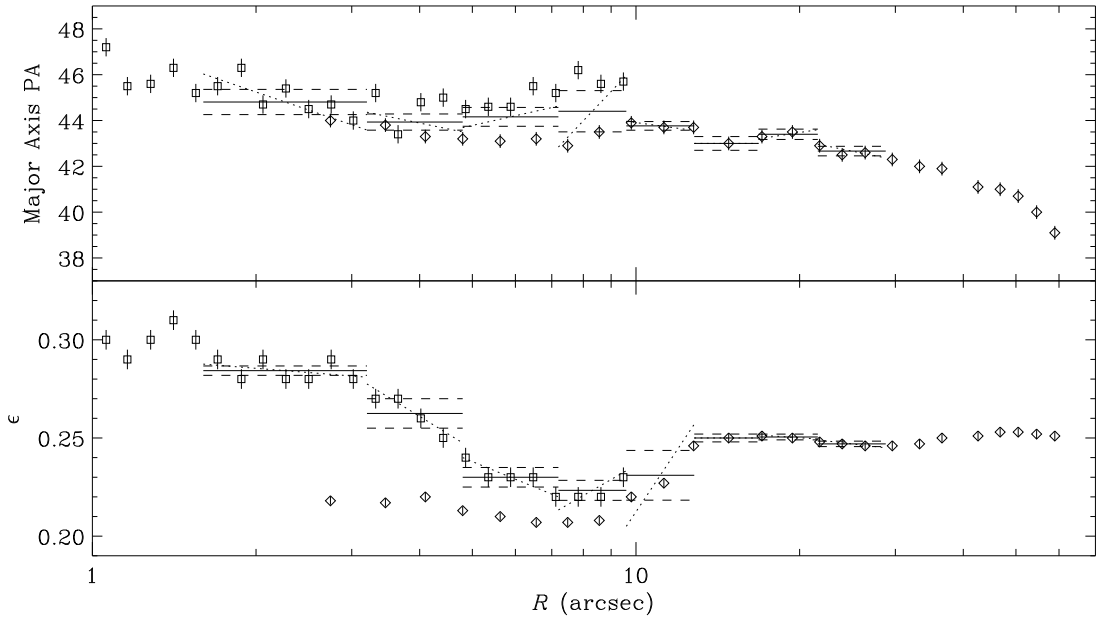}
\caption{Isophotal major-axis position angle ({\em top\/}) and ellipticity
({\em bottom\/}) plotted against mean radius. Data from \citet{Forbes} and
\citet{Benderetal88} are shown as squares and diamonds, respectively.
{\em Solid\/} and {\em dashed lines\/} show means and 1$\sigma$ errors,
resectively, over the radial bins used for model fitting. {\em
Dotted lines\/} indicate fits to the gradients over each bin, which contribute
to the errors as explained in \S\ \ref{s.data}.
\label{f.photometry}}
\end{figure*}

We re-bin the VF on to a coarser polar grid for comparison with the models.
This is purely a computational expedient, since CPU time and storage
requirements scale, respectively, with the number of angular and radial
bins. We are not losing information because there is no
significant structure in the VF at scales smaller than the grid. We set
the centres of the radial bins at $2\farcs 4$, $4\farcs 0$, $6\farcs 0$,
$8\farcs 4$, $11\farcs 2$, $14\farcs 8$, $19\farcs 2$, and $25\farcs 2$.
The central $1\farcs 6$ is excluded from the fit because the VF
projection is very inaccurate when the rotation curve is steeply rising.
Each radial bin is divided into between 10 and 24 approximately square
angular segments, which contain between 4 and 12 Voronoi nodes. The grid
is shown overplotted in Figure~\ref{f.data}a. We also antisymmetrize the
VF by folding and averaging points on opposite sides of the
centre. The symmetric component, i.e., the difference between the rebinned
data and the folded version, has an RMS amplitude of $3.3\kms$, which is
consistent with zero given the observational errors. We use the symmetrized
VF to compare with the models, since the latter are by construction
point-symmetric. The antisymmetrized, rebinned VF is shown in
Figure~\ref{f.data}b.

Profiles of isophotal ellipticity and position angle are
taken from the {\it HST\/}/PC F555W photometry of \citet{Forbes} and
ground-based $V$-band photometry of \citet{Benderetal88}. The profiles
are plotted in Figure~\ref{f.photometry}. Combining the datasets is
slightly complicated by the fact that neither set of profiles was
published with error bars; the error bars in the figure represent
a guess as to the quality of the data. We discard the ground-based ellipticity
data for $R<9.6\arcsec$, because it differs significantly from the
{\it HST\/} data in the sense expected from seeing effects. We average
the remaining data over each radial bin, and also fit for a gradient across
the bin. The adopted errors represent the formal error in the mean added
in quadrature with 19\% of the systematic variation across the bin.
The dotted black lines in Figure~\ref{f.data}b show the isophotes, plotted at
mean radii matching the radial bin centres.

\subsection{Fitting\label{s.fitting}}

Model fitting is done using Bayesian methods, which yield probability
distributions for the axis
ratios at each radius and the orientation of the system. The procedure
is conceptually straightforward: for each computed model, the likelihood
of obtaining the observed velocity field and isophotal ellipticity and
position angle profiles is calculated from the model prediction
and the known measurement errors. This is repeated for $\sim 10^8$
models covering the space of shape, orientation, and internal dynamical
parameters, creating a multidimensional likelihood function.
The likelihood is multiplied by a model for the parent distribution of
these parameters (the ``prior''), and integrated over the parameters that
one is not trying to constrain. The normalized result is the so-called
``posterior'' probability distribution for the parameters of interest.

\begin{figure*}
\includegraphics[width=6in]{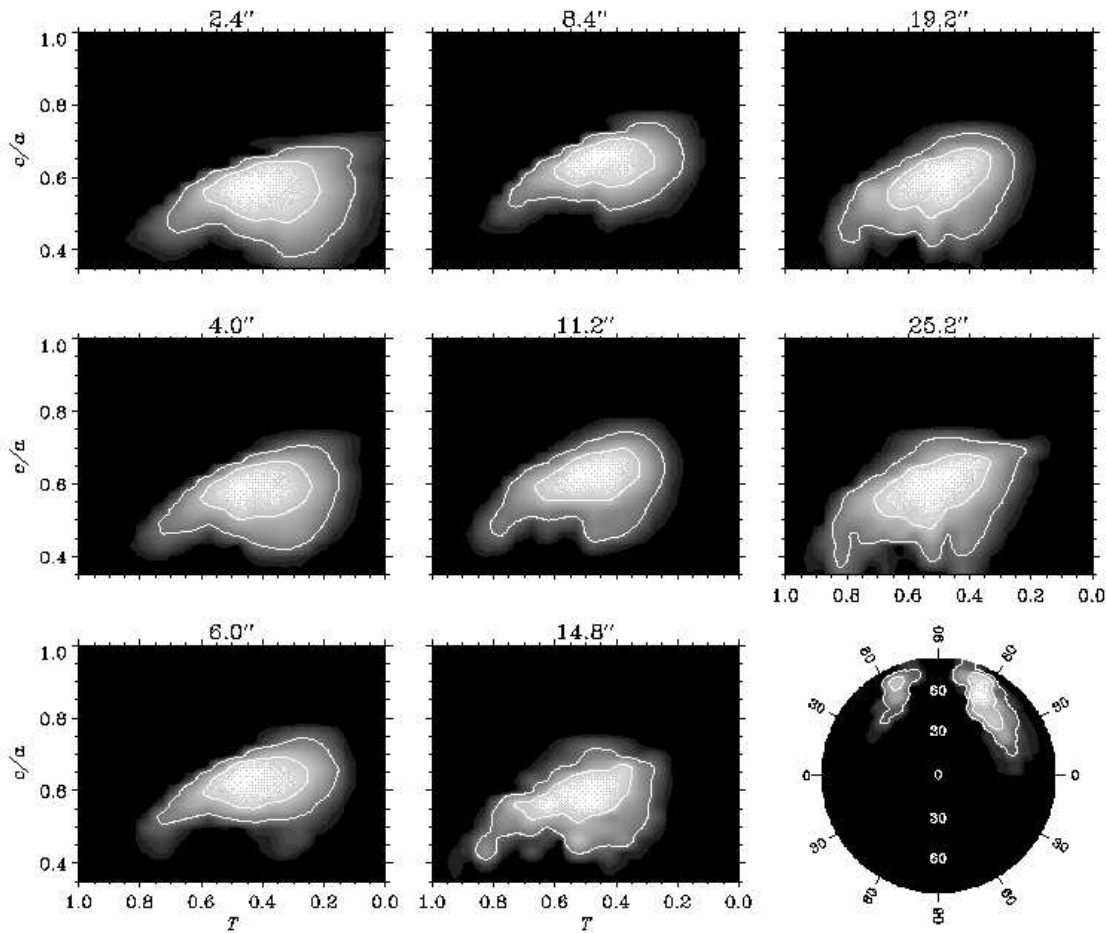}
\caption{Intrinsic shape profile and orientation of NGC~4365, based on the
``maximal ignorance'' unweighted average of all models and a flat parent
shape distribution. Rectangular panels
show probability densities, in the space of triaxiality $T$ and short-to-long
axis ratio $c/a$, for each radial zone; radii are indicated above each panel.
Round figure at lower right shows probability density for the line of sight
over a hemisphere, in Lambert equal-area polar projection;
centre, right and left edges, and top and bottom edges of the figure
correspond to views down the short, middle, and long axes of the galaxy,
respectively. The direction of the angular momentum vector corresponds to a
point along the upper half of a vertical line through the centre. The
greyscale is logarithmic in all panels, and white contours enclose 68\% and
95\% of the total probability.
\label{f.shape_flat}}
\end{figure*}

\begin{figure*}
\includegraphics[width=6in]{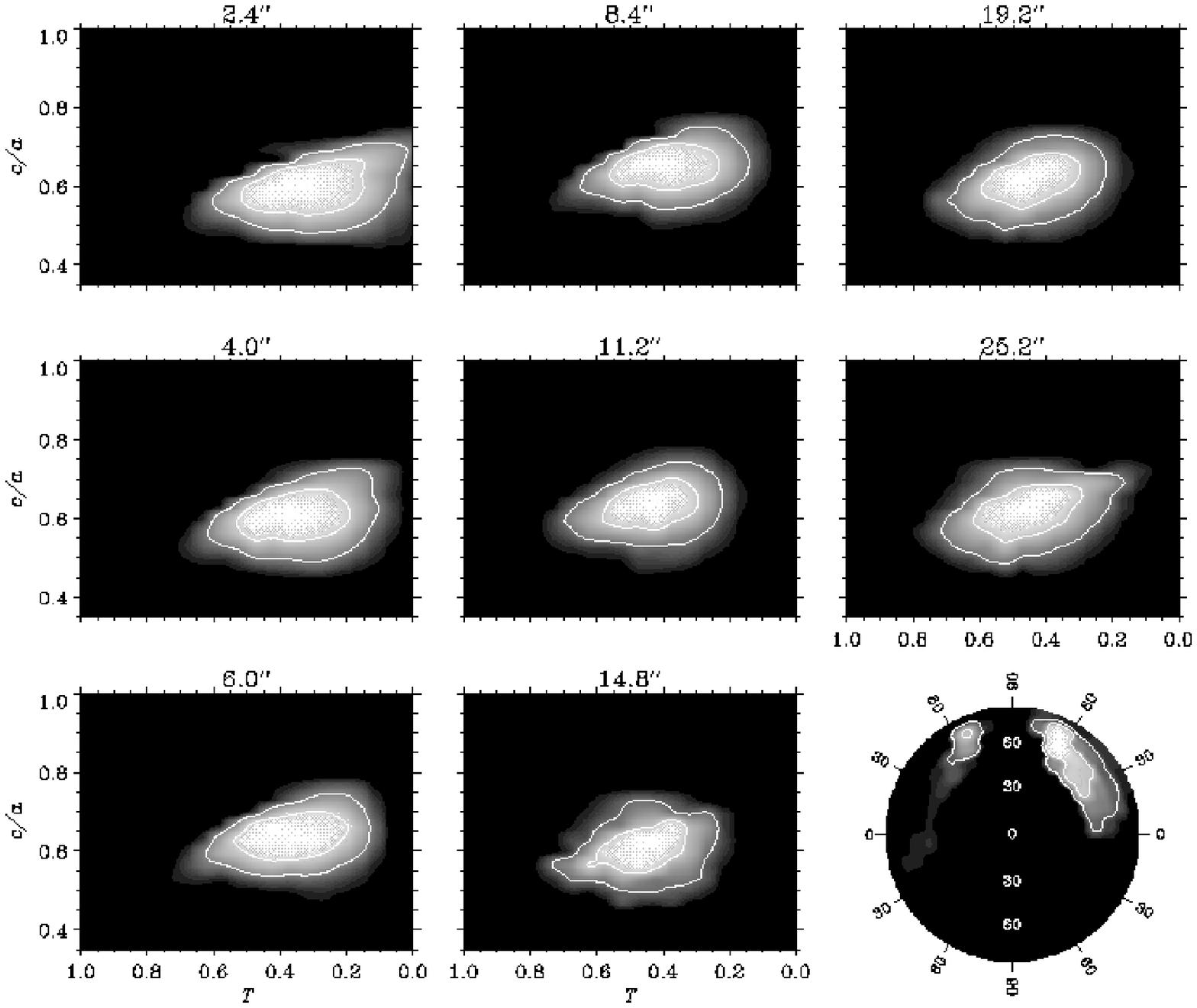}
\caption{Same as Fig.\ \protect{\ref{f.shape_flat}}, but using the parent
shape distribution from \citet{BS}.
\label{f.shape_bs}}
\end{figure*}

In practice, the procedure is slightly complicated by technicalities
needed to handle the 42-dimensional parameter space. The likelihoods for
each radial bin are stored on a $20 \times 20$ grid in triaxiality $T$
and axis ratio $c/a$, and an angular grid with approximately square bins
of $\sim 10\arcdeg$ resolution. The contribution to the likelihood from
the VF and ellipticity data can be calculated independently at each radius,
reducing storage demands considerably. We find it desirable at this
stage to introduce a penalty for configurations that require
unrealistically high internal velocities because the line of sight is
close to the rotation axis (see the Appendix for a full explanation).
Turning the likelihoods into the shape profile of the galaxy formally
involves the construction of a joint distribution in 16 variables. To make
this more manageable, we compute the projections of this distribution
into each of the shape planes, and use these ``marginal distributions''
as the basic description of the shape profile. The procedure for combining
the likelihood fits at each radius and including the isophotal PA data is
explained in detail in the Appendix of \cite{SDS}.

Construction of the prior involves a few subtle points which are described
in detail in the Appendix of this paper. In brief, the prior
distribution is assumed to separate into four factors:
\begin{eqnarray}
F_{\rm prior} &=& {1 \over 4 \pi} F_{\rm par}[T_i,(c/a)_i]
F_{\rm cor}[T_i-T_j,(c/a)_i-(c/a)_j] \nonumber\\
&&\times F_{\rm int} [C_i, k_{Xi}, k_{Zi}; T_i,(c/a)_i].
\end{eqnarray}
The leading factor is the
angular part, which we take to be isotropic over the viewing sphere.
The second is the parent shape distribution; in principle this
is a model for the shape distribution of all
ellipticals, although for modeling a single system
it is arguably better to use a flat distribution. We will show results
for both a flat distribution and for the parent distribution derived from
published long-slit data \citep{BS}. The third part describes
correlations in the joint shape distribution among the radial bins.
These correlations reflect the fact that we expect the intrinsic shape
to change continuously with radius. The last factor describes the
distribution of the internal parameters, $C$, $k_X$, and $k_Z$. In the
``maximal ignorance'' estimate we assume no correlations of kinematics with
radius. In other words, we explicitly allow for the possibility that there
can be kinematically different populations dominating at different radii,
bu we make no attempt to distinguish these populations, beyond their effects
on the local mean streaming motions. In our standard model grid we also
include a subset of models in which $C$ is correlated with shape at each
radius, but these models give results consistent with the others.

\section{Results\label{S.RESULTS}}

\subsection{Intrinsic Shape Profile\label{s.shape}}

The basic results are shown in Figure~\ref{f.shape_flat}. The 
rectangular panels show the posterior probability distributions for the
shape of the isodensity surfaces in the eight radial bins, based on an
unweighted average of all computed models. Each panel
depicts the $(T,c/a)$ plane and is plotted so that oblate spheroids
($T=0$) lie along the right edge and prolate spheroids ($T=1$) lie along
the left edge. Spheres occupy the entire
top margin at $c/a=1$, and the bottom margin, at $c/a=0.35$, corresponds
to the observed upper limit on E galaxy ellipticities. The white curves
indicate the 68\% and 95\% highest posterior density (HPD) regions, i.e.,
the contours enclosing 68\% and 95\% of the total probability, and may
be thought of as $1\sigma$ and $2\sigma$ error regions.

It is clear that NGC~4365 is an almost maximally triaxial system. The
preferred triaxialities are in the vicinity of $T \approx 0.4$--$0.5$, and
axisymmetric and near-axisymmetric shapes are strongly ruled out, except
perhaps in the innermost bin. There is an indication of an outward
triaxiality gradient. The galaxy is also quite a bit flatter than
it appears; photometry alone requires only that $c/a \lesssim 0.75$.

The results in Figure~\ref{f.shape_flat} are for a flat prior shape
distribution, which is appropriate for one system modeled in
isolation. But this part of the prior actually represents the parent shape
distribution for the full population of ellipticals, which can be
estimated from published long-slit kinematics. \citet{BS} model the
\citet{DB} sample of 13 ellipticals, using methods very close to those
of this paper, and derive several different parent distributions based
on different assumptions for the dynamical prior. Their ``maximal
ignorance'' result (their Fig.\ 2a) corresponds most closely to the
dynamical prior used here. This distribution has broad peaks centred
on the oblate and prolate limits, approximately at $(T,c/a)=(0,0.68)$
and $(1,0.82)$, with a broad ridge between them. Spheres and very
flat, prolate-triaxial shapes have very low amplitude.

The shape profile of NGC~4365 using the \citet{BS} parent distribution is
shown in Figure~\ref{f.shape_bs}. The results are similar to those using
the flat parent distribution, differing primarily in that the extended
tail toward large $T$ and small $c/a$ is lacking. It is worth noticing
that the single-galaxy shape constraint is improved when information
from other objects is included in the form of the parent distribution.
One can expect that the greatly superior parent distribution that will
be derivable from the SAURON sample will permit even more precise
single-galaxy shapes to be determined.

\begin{figure}
\includegraphics[width=3.3in]{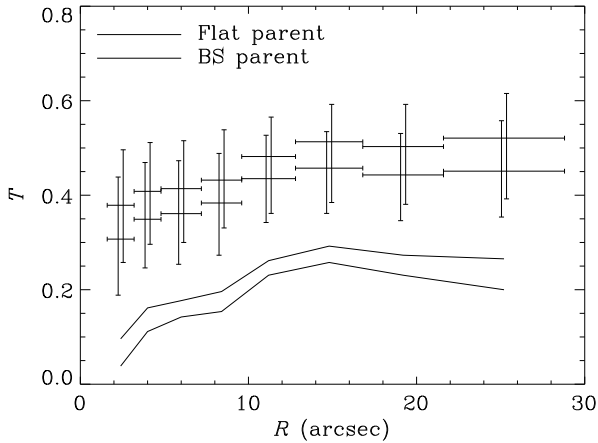}
\caption{Radial triaxiality profiles derived from Figures
\protect{\ref{f.shape_flat}} and \protect{\ref{f.shape_bs}} for the flat
and \citet{BS} parent distributions. Points with error bars show expectation
values and 68\% HPD intervals ($1\sigma$ error bars); horizontal bars
indicate the widths of the bins. Solid lines below denote 99\%-confidence
lower limits on $T$.
\label{f.tprofile}}
\end{figure}

One dimensional triaxiality profiles can be straightforwardly
calculated by integrating the distributions in Figures
\ref{f.shape_flat} and \ref{f.shape_bs} over $c/a$. The results are
shown in Figure~\ref{f.tprofile}. Error bars show the 68\% HPD
intervals, which are nearly centred on the expectation values. The
systematic effect of changing the parent shape distribution is smaller
than the statistical errors. There is an indication of a weak
triaxiality gradient, but constant $T$ cannot be excluded.
Possibly of greater interest for constraining
the effects of chaos are lower limits on $T$; the continuous curves in
the figure show 99\%-confidence limits, which are significantly
non-zero except possibly in the innermost bin.

\subsection{Orientation\label{s.orientation}}

\begin{figure}
\includegraphics[width=3.3in]{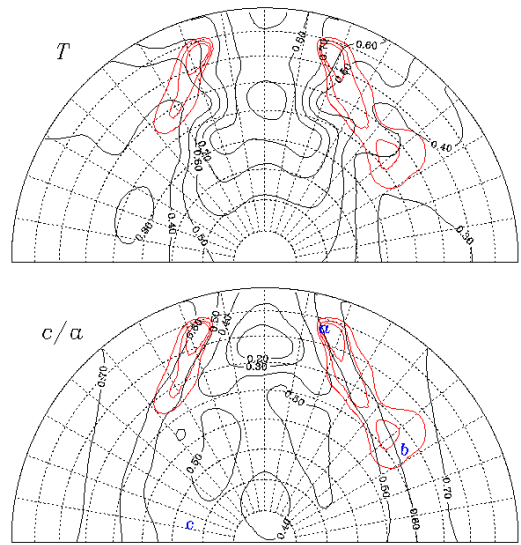}
\caption{Lambert equal-area projections of half of the viewing hemisphere,
showing ({\em top\/}) $T$ and ({\em bottom\/}) $c/a$ values of the best-fitting
optimized model at each orientation ({\em black contours\/}). Overplotted
in red are contours of $\Delta \chi_{\rm pen}^2$, at levels of $2.3$, $6.17$,
and $11.8$. Models outside the outermost red contour are formally excluded at
$>99\%$ confidence. {\em Dotted lines\/} show an angular grid at
$10\arcdeg$ spacing for reference. Letters in the lower panel indicate
the 3 models shown in detail in Fig.\ \ref{f.models}.
\label{f.optimized}}
\end{figure}

The round panels in the lower right of Figures \ref{f.shape_flat} and
\ref{f.shape_bs} show the posterior distributions for the line of
sight, over one hemisphere.\footnote{Views from antipodal points on the
sphere are mirror images of each other. The likelihoods recorded in each
angular bin are averages of the two antipodal views. The ambiguity is
trivially resolved {\em a posteriori.}}\ These are the orthogonal
projections of the full posterior probability, integrated over shape,
in contrast to the shape distributions which are integrated over orientation.

The line of sight is constrained to two fairly narrow strips, in the quadrant
containing the angular momentum vector. That is, the true rotation axis
in the outer part of the galaxy is most likely pointing in the general
direction of the observer, as opposed to lying in the plane of the sky.
Lines of sight $\sim 30\arcdeg$ from the long axis are preferred, and
those near the short axis are strongly excluded. These constraints are
only slightly affected by the choice of the parent shape distribution.

The two strips in the orientation figures correspond to views on opposite
sides of the $(x,z)$ plane, producing the same velocity field.
Mirror-image points on the left and right sides of the figure differ
only photometrically: for a given triaxiality gradient the isophotes twist
in the opposite direction. If there were no observed isophotal twist, the
figure would be left-right symmetric. The small asymmetry in the figure
reflects the small ($\sim 2\arcdeg$) observed twist over the fitted region.

\subsection{Goodness of Fit; $\Delta\chi^2$ Confidence Limits\label{s.chi}}

\begin{figure*}
\includegraphics[width=6in]{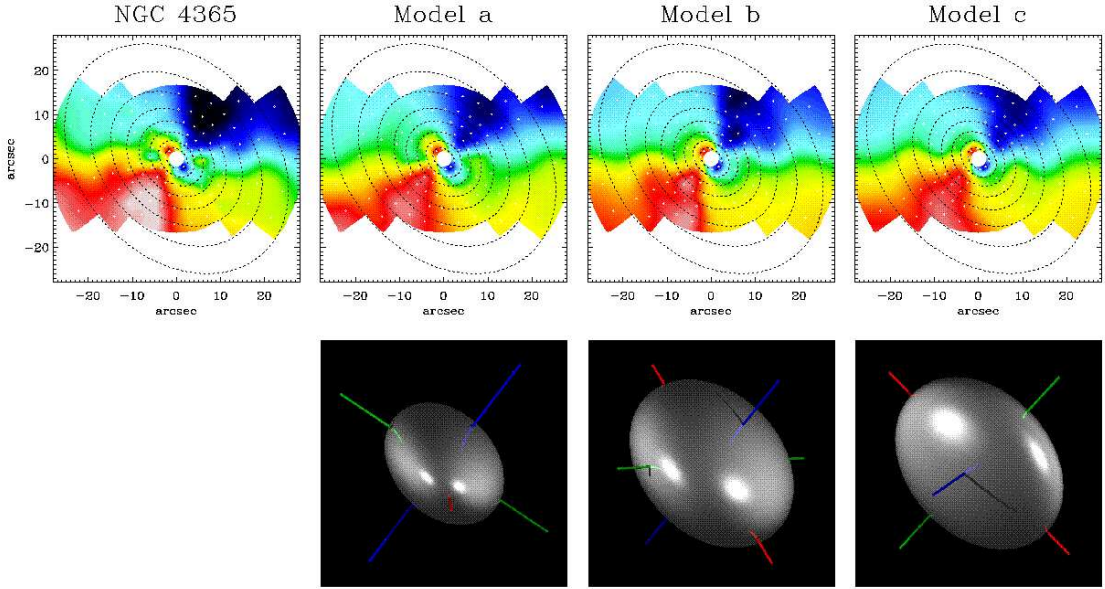}
\caption{Data ({\it top left\/}) and models in the observational
domain. {\em Top row\/}: isophotes and velocity fields, plotted as in
Fig.\ \ref{f.data}. {\em White points\/} mark bin centres, where the fit
is performed. {\rm Bottom row\/}:
ray-traced images of the mean shapes, as oriented in
the sky. The shapes are rendered as polished metallic ellipsoids,
symmetrically illuminated by lights to the right and left behind the
observer.  Red, green, and blue rods indicate the long, middle, and
short axes. Models $a$, $b$, $c$ correspond to those indicated in Fig.\
\ref{f.optimized}, with ($a$) good, ($b$) marginal, and ($c$) poor fits.
\label{f.models}}
\end{figure*}

A disadvantage of the Bayesian approach is that a posterior probability can
be calculated even if the fit to the data is poor. It is therefore necessary
to check separately that the best models are actually adequate fits in
a $\chi^2$ sense. We couple our modeling code to a standard multidimensional
optimizer and calculate, for each angular bin centre, the combination of
shape and dynamical parameters that minimizes $\chi^2$ at that orientation.
This allows us also to verify whether the HPD regions agree with confidence
limits determined from $\chi^2$ contours.

We have 81 data points (66 velocities, 8 ellipticities, and 7 PA
differences), and, at fixed orientation, 40 parameters ($T$, $c/a$,
$C$, $k_X$, and $k_z$ in each radial bin).
We minimize a $\chi^2$ statistic that includes the same penalties (for
shape differences between adjacent bins and large internal $v/\sigma$)
that are used in the Bayesian treatment. This adds 15 constraints,
making $\nu_{\rm pen}=56$ degrees of freedom. The best optimized model has
$\chi_{\rm pen}^2/\nu = 1.14$. This is an acceptable fit; the
probability of a larger $\chi_{\rm pen}^2$ occurring by chance is $0.22$.
We also calculate, but do not optimize, the unpenalized
$\chi_{\rm un}^2$ measuring only the fit to the data. The model
optimized on $\chi_{\rm pen}^2$ has $\chi_{\rm un}^2/\nu_{\rm un} = 1.41$,
indicating an adequate but not spectacular fit.

In Figure~\ref{f.optimized} we plot in black the mean shape (unweighted average
over the 8 radial bins) of the best optimized model at each orientation, for
the half of the viewing hemisphere containing the viable models. These contours
indicate only the best model for each line of sight, not whether these
models are statistically good fits. Overplotted in red are contours of
$\Delta \chi_{\rm pen}^2$, at levels of $2.3$, $6.17$, and $11.8$ above the
minimum. Formally these correspond to 68\%, 95\%, and $99.7\%$ confidence
intervals for two degrees of freedom, so lines of sight outside the last
red contour are strongly ruled out. The models inside the red contours, which
are not ruled out,
have triaxialities $T$ between $0.32$ and $0.72$, and flattenings $c/a$
between $0.40$ and $0.68$. This is consistent with the results in
Figure~\ref{f.shape_flat},
which include many more models than just the best ones
on each line of sight. Furthermore, the correspondence between the red
$\Delta \chi_{\rm pen}^2$ contours in Figure~\ref{f.optimized} and the
HPD contours in the bottom right of
Figure~\ref{f.shape_flat} is qualitatively good; they clearly
identify the same preferred lines of sight. The agreement is not exact,
however. We attribute this to the difference between $\Delta \chi^2$ values
computed, on the one hand, along an optimised $\chi^2$ valley floor in the
space of the non-orientation parameters and, on the other, averaged over the
whole valley at quite coarse resolution.

Notice that, on the viewing hemisphere, the $\chi^2$ valley is nearly
parallel to the contours of
optimized $c/a$, but cuts across the contours of optimized $T$. The fact
that triaxialities $\sim 0.45$ are optimal over the broadest range of
orientations is consistent with those values of $T$ being most probable
(Figs.\ \ref{f.shape_flat}, \ref{f.shape_bs}, and \ref{f.tprofile}).
One should not place too much emphasis on the exact position of the
$\chi^2$ minimum. Visual inspection of the acceptable models shows that
different models fit different parts of the data better than others, and
an alternative rebinning of the VF might easily have moved the
$\chi^2$ minimum by $\sim 10\arcdeg$ along the valley.

The blue letters in Figure~\ref{f.optimized} identify three representative
models which we show in detail in Figure~\ref{f.models}. Models $a$,
$b$, and $c$ correspond to good, marginal, and bad fits respectively,
having $\chi_{\rm un}^2/\nu$ values of $1.4$, $1.7$, and $2.8$, and
$\chi_{\rm pen}^2/\nu$ values of $1.1$, $1.3$, and $2.3$. The top
row of Figure~\ref{f.models} shows the projected isophotes and velocity
field of each model, with the observed data from Figure~\ref{f.data}b reproduced
at top left. The
bottom row shows a rendered-surface representation of an ellipsoid
with the mean shape of each model, at the correct orientation. Clearly the
acceptable models are able to reproduce the detailed kinematic and photometric
structure of the galaxy. One can see from the figure that the biggest
difficulty is simultaneously fitting the rotation amplitude
and the steep velocity gradient near PAs $100\arcdeg$ and $280\arcdeg$. 
The unacceptable model $c$ does a particularly poor job of reproducing this
gradient, and its kinematically distinct core region is too small.

\subsection{Orbital Structure\label{s.orbits}}

Because we have allowed the orbit populations, by way of the velocity
boundary condition, to vary freely with radius, the inferred shape profile
and orientation of the system are essentially free of any assumptions
about dynamical subsystems in the galaxy or its formation history. But, by
the same token, we are unable to constrain the details of the stellar
populations with these coarse models. That would be a job for much more
sophisticated modeling, making use, at least, of the full line-of-sight
velocity distributions, and possibly line-strength indices as well.
None the less, we can draw some conclusions concerning what
orbits dominate the net streaming motions in the inner and outer parts
of the galaxy.

\begin{figure*}
\includegraphics[width=6in]{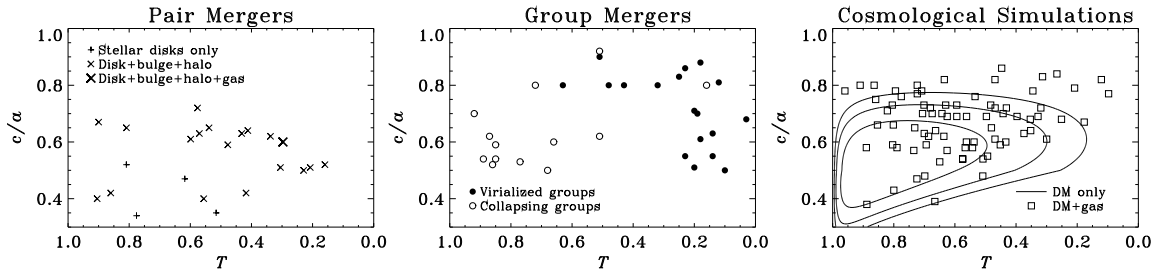}
\caption{Intrinsic axis ratios of merger remnants and dark haloes from
the literature. Each panel shows mean shapes or local shapes near the
half-mass radius, as reported by the authors.
{\em Left\/}: spiral-spiral merger remnants, without (Hernquist 1992, 1993,
Barnes 1992) and with (Barnes \& Hernquist 1996) gas dissipation.
{\rm Center\/}: Remnants of dissipationless group mergers (Weil \&
Hernquist 1996 as reanalysed by Statler et al.\ 2001, Zhang et al.\ 2002).
{\em Right\/}: Distribution of zero-redshift halo shapes in a dissipationless
$\Lambda$CDM
cosmological simulation (Jing \& Suto 2002) ({\em contours\/}; spaced by
factor of 2 in density) and individual shapes of luminous objects in
dissipational simulations (Sugarman et al.\ 2000, Meza et al.\ 2003).
\label{f.simulations}}
\end{figure*}

To explore the dynamics of the kinematically distinct core, we compare
three subsets of models in which the innermost zone is restricted
to certain values of the contrast $C$: a Z-tube dominated
configuration ($C=\infty,2$), an X-tube dominated configuration
($C=0,0.0625$), and a neutral configuration ($C=0.25,0.5$)\footnote{Z
tubes still dominate the total velocity field when $C=1$, as can be seen in
Fig.\ \ref{f.streamlines}.}\ We compute the maximum of
the joint likelihood $L(T,c/a,\phi,\theta)$ in each radial bin,
integrated over dynamical parameters but not over orientation. We find
the Z-dominated configuration to have the highest peak likelihoods, with
those in the neutral and X-dominated configurations smaller by factors
of 5 and $3\times 10^4$, respectively. Insisting that the core rotation
is entirely due to Z-tubes barely alters the shape profile. Axisymmetry
in the core is still excluded at a $>2\sigma$ level. Photometrically,
the isophotes between $1\arcsec$ and $3\arcsec$ are slightly discy
\citep{Carollo,Rest}, and the large $h_3$ values in the same region are
suggestive of discy kinematics. But a stellar disc embedded in a larger
triaxial system will not, in general, be axisymmetric.

We can also investigate the dynamics of the
region outside the core, by following the same procedure and
restricting $C$ in the outer 4 zones. Here we find the situation reversed:
the X-dominated configuration has the largest peak likelihoods, with the
neutral and Z-dominated configurations down by factors of 4 and $10^6$,
respectively.

We conclude, therefore, that the mean motions are most likely dominated
by long-axis tubes in the main body of the galaxy and by short-axis tubes
in the kinematically distinct core; this is essentially the configuration
advocated by \citet{Surma} for NGC~4365, and by \cite{FIH} for the
similar system NGC~4406. It is virtually certain that the main
body is not short-axis tube dominated, and that the core is not long-axis
tube dominated. But configurations in which both tube families contribute
significantly all through the galaxy cannot be ruled out. In fact, all
of the models in Figure~\ref{f.models} require Z tubes in the main body
to move the apparent
rotation axis away from the position angle of the projected long axis.

\section{Discussion\label{S.DISCUSSION}}

\subsection{The Importance of Quantitative Measurements of Galaxy Shape}

The distribution of
three-dimensional shapes is an under-utilized, but potentially
powerful diagnostic of galaxy formation physics. Numerical simulations of
structure formation and galaxy mergers can now predict the axis ratios
of the zero-redshift haloes or final merger remnants, and the results depend
on the processes involved. Figure~\ref{f.simulations} shows representative
results from the literature. Pair mergers of pure discs produce very
flat, very triaxial remnants \citep{Hern92}. Including
dense bulges or moderate gas dissipation tends to make the final systems rounder
and less prolate \citep{Barnes92,Hern93,BH96}, in the former case because
the merging systems are themselves rounder, and in the latter because of the
scattering of low angular momentum orbits by the central mass accumulation.
Significant dissipation in
gas-rich mergers can flatten the final system if the gas settles into a
disk and subsequently forms stars \citep{HB91,Barnes02,Lamb}.
Pure dark matter haloes formed in cosmological simulations are also very
flat and prolate \citep{JingSato}; as in disc mergers, including some
dissipation tends to make the final systems rounder, though still strongly
triaxial \citep{Sugarman,Meza}. Most prominently,
mergers of groups should produce highly prolate or
oblate remnants, depending on whether the initial system is collapsing
or virialized \citep{Weil,Garijo,Zhang}.
Clearly there is discriminating power in the shape distribution.

The SAURON instrument has produced stellar kinematic maps for two dozen
elliptical galaxies \citep{EmsellemSauron}. This paper constitutes the
first step in using these data to determine the shape distribution.
The results of the preceding section illustrate the power of
two-dimensional data to constrain the shapes of individual systems.
The axis ratios of NGC~4365 are determined to within approximately $0.1$,
at the $1\sigma$ level. We have also modeled the SAURON VF for NGC~3379,
and find that we can constrain the shape to comparable precision. The
NGC~3379 results are very similar to those obtained from 7 long-slit profiles
by \cite{Statler3379}. This is not surprising, since the VF is very
simple and symmetric, and the SAURON map does not contain much additional
information beyond what could be interpolated from the long-slit data.
One may justifiably anticipate that it will be possible to determine the
shapes and orientations of other objects in the SAURON sample to comparable
precision using these methods, to estimate the parent
shape distribution for the sample with the approach of \citet{BS}, and
to study the dependence of this distribution on galaxy luminosity or
environment. These topics will be the focus of future papers.

\subsection{Effects of Chaos; Limits on Central Black Hole Mass}

NGC~4365 is an old system. From the metal and Balmer line indices
\citet{DaviesSauron} infer an age $\gtrsim 12\gyr$ for the bulk of
the stellar population, with at most a few percent admixture of a
younger ($5\gyr$) population in the centre. Absolute stellar
ages are uncertain, however, and recent photometric \citep{Puzia} and
spectroscopic \citep{Larsen} studies of globular clusters indicate the
presence of an intermediate-age population. A minor merger $2$--$5\gyr$
ago is not unlikely, although a significant event would probably have
left photometric traces at large radii, which are not seen \citep{DaviesSauron}.

None the less, {\em any\/} age in the range of several Gyr implies that the
system is dynamically very old. We can define a local
relative dynamical age $\tau_d(r)$ as the ratio of the assembly
age $t_\star$ to the crossing time at projected radius $r$. We 
approximate the latter as
$r/[v_m^2(r)+\sigma^2(r)]^{1/2}$, where $v_m(r)$ is the maximum mean
velocity at $r$. The result can be represented extremely well by the
relation
\beq
\tau_d \approx 3\times 10^4 \left({t_\star \over 10\gyr}\right)
	 \left({r \over 1\arcsec}\right)^{-1.1};
\eeq
thus the entire SAURON field is hundreds of dynamical times old even if
the galaxy was assembled as recently as $2\gyr$ ago.

The long-term survival of triaxiality in systems with central density
cusps or black holes has been questioned many times \citep[and
references therein]{LakeNorman,Gerhard85,MerrittReview}. Central mass
concentrations
can render a significant fraction of the box orbit phase space chaotic.
Since box orbits are essential for maintaining triaxiality, it is
believed that growth of a cusp or black hole can force a triaxial system
to evolve toward axisymmetry or sphericity. That this has clearly not
happened in NGC~4365 allows us to limit the maximum mass of a
central black hole.

Calculating a precise upper limit is not yet possible.
Surprisingly few numerical studies of chaos-driven morphological
evolution have been performed, and these are not entirely appropriate to
the properties of NGC~4365. \citet{MQ} start with a prolate-triaxial
($T\approx 0.6$, $c/a \approx 0.5$) equilibrium system with a
flat, constant-density core, and grow central dark masses amounting to
$0.3\%$, $1\%$, and $3\%$ of the galaxy mass. They see a large, rapid
decrease in $T$ that effectively axisymmetrizes the entire system for
mass ratios $\geq 1\%$. Below this mass the change is less drastic, but
still pronounced. They also find that the shape evolution
continues even after the central object reaches its final mass.
\citet{Sellwood} argues that this late evolution is a numerical artefact,
but confirms Merritt \& Quinlan's basic result that mass ratios larger than
1\% globally destroy triaxiality. \citet{HB} perform a similar experiment
in which a 1\%-mass black hole is grown in a somewhat rounder ($T=0.54$,
$c=0.70$) triaxial system with an initial $r^{-1}$ density cusp.
Unlike the flat-core systems, the cuspy system does not experience a
global destruction of triaxiality. Its inner regions tend toward a
spherical shape at approximately constant $T$, while the outer half of
the system remains flattened and triaxial. The black hole creates a
spherical region around itself containing several times its own mass
in stars. The interpretation of these simulations is further complicated
by the existence of flattened triaxial equilibria containing both cusps and
central masses. \citet{Poon2} construct equilibrium models for triaxial
($T=c/a=0.5$) nuclei with $r^{-1}$ and $r^{-2}$ cusps by Schwarzschild's
method. They confirm using $N$-body simulations that the shapes, even
down to radii enclosing only a few times the black hole mass,
are stable for $\sim 10$ crossing times. It is not known
whether they would persist over tens of thousands of crossing times, as
NGC~4365 would require. For whatever reason, these equilibria seem not
to be found by the simulations that begin with no black hole. Evidently they
do not have strong basins of attraction around them when the central mass
grows dynamically.

We accept the basic result from \citet{MQ}, \citet{HB}, and
\citet{Sellwood}, that a black hole $>1\%$ of the system mass will drive
a rapid evolution of a region whose size is a significant fraction
of the effective radius toward either axisymmetry or sphericity. This can
be decisively ruled out in NGC~4365. To estimate the system mass, we use
the total $V$-band magnitude,
$V_T^0=9.5$, given in the RC3 \citep{RC3}, a distance modulus of $31.55$
\citep{TonrySBF}, and a mass-to-light ratio $M/L_V=6$ obtained by fitting
an isotropic oblate Jeans model to the full SAURON data set. This $M/L_V$
is consistent with the results of \citet{GebhardtModels}, who find a mean
$M/L_V=5.3$ for 11 ellipticals, with a dispersion of $2.1$.
We obtain $M=3.3 \times 10^{11} \msun$ for NGC~4365.
This implies that a black hole of mass $M_{\rm BH} > 3
\times 10^9 \msun$ would either globally axisymmetrize the galaxy or at
least render the inner $10^{10}\msun$ spherical. By integrating the
$\gamma=0.13$ Nuker model fit to the surface brightness profile \citep{Rest},
we find
that this mass is enclosed in projection by the isophote with a mean radius of
$3\arcsec$. Although we see indications of declining triaxiality at
these radii, there is no significant rounding of the isophotes, which
would be seen if the inner regions were becoming spherical. We conclude
that the long-lived triaxiality in NGC~4365 rules out black holes of
$M_{\rm BH}>3 \times 10^9 \msun$.

There is, as yet, no direct evidence for a black hole in NGC~4365. But
for a mean dispersion of $260\kms$, the $M_{\rm BH}$-$\sigma$ relation
\citep{NukerMsigma,FM,TremaineMsigma} would predict $M_{\rm BH}\approx
4 \times 10^8\msun$. This corresponds to $\sim 0.1\%$ of the system
mass, and probably would not drive a global shape change. If the
single simulation of \citet{HB} can be extrapolated to lower masses,
then a black hole of this mass might be expected to affect the isophotes
in the inner $0\farcs 9$. In the simulation, an $r^{-2}$ density
cusp grows within the classical radius of influence of the black hole,
which, for NGC~4365's central dispersion of $270\kms$, would subtend
$0\farcs 1$. {\it HST\/}/WFPC2 surface photometry \citep{Rest,Carollo}
does show a gradual rounding of the isophotes interior to $1\arcsec$,
reaching a minimum ellipticity $\sim 0.08$ at about $0\farcs 16$.
The central photometric structure is complicated, and may be affected by
dust. There is no steepening of the $R$-band brightness profile within
$0.1\arcsec$, but there does seem to be a central, unresolved source
that is slightly ($\sim 0.02$ mag) bluer than its surroundings \citep{Carollo}.

Thus, while the persistent triaxiality of NGC~4365 excludes large black
holes, a more modest object, on the scale predicted by the
$M_{\rm BH}$-$\sigma$ relation, is probably small enough not to globally
alter the shape of the galaxy and can be consistent with the data.

\subsection{Figure Rotation}

We have shown in \S\ \ref{s.orbits} that net streaming motions in
long-axis tubes contributes significantly to, and probably dominates,
the observed velocity field. The direct and retrograde branches of these
orbits generate the same spatial density distribution if there is no
rotation of the potential. However, if the galaxy is tumbling about the
short axis, these orbits acquire a tilt with respect to the symmetry
plane. This tilt increases for larger orbits, and is in the opposite
direction for two senses of circulation. Since there must be an
imbalance in the populations of the direct and retrograde branches,
figure rotation would induce an intrinsic twist in the galaxy
\citep{SchRot}. The small
observed photometric twist allows us to limit the tumbling rate.

The tilt of the periodic planar orbits out of the $(y,z)$ plane caused
by figure rotation about the $z$ axis was examined by \citet{vAKS} and
\citet{HMS}. These so-called ``anomalous'' orbits are the parents of the
long-axis tubes \citep{SchRot},
which we will assume acquire roughly the same tilt as
the parent of the same energy. This assumption is by no means firmly
established; the behavior of general orbits in tumbling figures has been
studied only superficially (Salow 1998, unpublished). \cite{vAKS}
show that the tilt angle $\alpha$ can be approximated by
\beq
\tan\alpha = {2 \over 1 - (k/\omega)^2 + (\Omega/\omega)^2}
	\cdot {\Omega \over \omega},
\eeq
where $\omega$ and $k$ are the frequencies of the periodic motion in the
plane and of oscillations out of the plane, respectively, in
the absence of figure rotation, and $\Omega$ is the tumbling
frequency. For slow tumbling, $\Omega$ can be ignored in the first
factor. The ratio $k/\omega$ is a function of the axis ratios, and
should be smaller than 1 for orbits around the long axis. For a
triaxial modified Hubble model with $T=0.8$ and $c/a=0.5$, \cite{vAKS}
calculate $k/\omega \approx 0.65$. Our models for NGC~4365
suggest shapes not quite as extreme as this. Let us
suppose that $k/\omega \approx 0.8$, which is appropriate for a prolate
logarithmic potential with a density axis ratio of roughly $0.6$.
For NGC~4365 we approximate
$\omega$ by $2^{1/2} \sigma/r$, with $\sigma \approx 220\kms$ the
one-dimensional dispersion at $R \approx 20\arcsec$, and find that
\beq
\alpha \approx 0.4\arcdeg \left({P_{\Omega} \over 1\gyr}\right)^{-1}
	\left( {r \over 1\arcsec} \right),
\eeq
where $P_{\Omega} = 2\pi/\Omega$ is the tumbling period.

It is interesting that the tilt is predicted to be linear in $r$.
NGC~4365, in fact, has a linear isophotal twist of $0.09\arcdeg/\arcsec$
for $2\arcsec < R < 60\arcsec$. How the anomalous orbit tilt translates
into an observed isophotal twist is extremely complicated, depending on
the exact orbit populations, axis ratios, and viewing geometry, and is
beyond our ability to predict in detail. However, if we imagine that the
observed twist and the anomalous orbit tilt are of the same order of
magnitude, then the observed twist could be caused by figure rotation
if the tumbling period $P_\Omega \sim 5\gyr$. Alternatively, we can
say that $P_\Omega \lesssim 500\myr$ is probably ruled out by the
absence of a larger twist. This tumbling rate would place corotation at
about 8 effective radii assuming a standard isothermal dark halo. It
seems, therefore, safe to conclude that figure rotation is dynamically
unimportant in the observable part of the galaxy.

If the isophotal twist is indeed due to figure rotation, then it is
worth asking in what direction the figure is rotating. \citet{vAKS}
performed this exercise for Centaurus A, in an effort to explain the
warp of the dust lane. They predicted that the southwest side of the
figure should be approaching and the northwest receding; unfortunately this
was later found to be opposite to the sense of rotation of the stars.
For NGC~4365, the required sense of figure rotation is northeast
approaching and southwest receding for models viewed as in
Figure~\ref{f.models}a and \ref{f.models}b. Short-axis tubes are needed in
these models, to displace the zero-velocity contour counter-clockwise on the
sky. The sense of rotation of the short-axis tubes is therefore in the same
direction as the figure tumbling, but opposite to the orbits of the same
family that dominate the decoupled core.

\subsection{Implications for Other Modeling Approaches}

Although the VF fitting method can constrain the axis ratios and the
orientation of the mass distribution, it does not make use of all of the
available kinematic data, and consequently cannot truly constrain the
full density profile. To do this, more sophisticated approaches are needed.

Jeans models have a long and distinguished history in galaxy dynamics.
These models are built on the Jeans equations \citep{Jeans,Chandra,BT},
which relate the
second moments of the velocity distribution to the density distribution
and the potential. Second-moment models are intuitively appealing, since
luminous ellipticals are supported by velocity anisotropy \citep{Binney78}.
But fully anisotropic models are not easily computed because, in general,
there are more independent second moments than there are Jeans equations
to constrain them. Very recently, an analytic solution to the triaxial
anisotropic Jeans equations has been demonstrated for systems with St\"ackel
potentials \citep{vdVJeans}. It may be possible to use this result to
create simple, approximate second-moment models, and to merge the
VF fitting and Jeans techniques (van de Ven et al., in preparation).

A full understanding of the dynamical structure of a galaxy requires a model
for the complete phase space distribution function. At present the only
workable technique to build such a model is Schwarzschild's method
\citep{Sch}. In this technique a finite library of test-particle orbits
is populated to reproduce a desired mass distribution and fit the photometric
and kinematic data. The newest Schwarzschild code for triaxial systems is
presented by \citet{VerolmeThesis} and \citet{Verolme}; but unfortunately,
no direct comparison of Schwarzschild and VF fitting results for NGC~4365
is yet possible. Schwarzschild models, by allowing any population of stars
on any orbit, have far more freedom than the continuity-equation models used
in VF fitting. We suspect that a Schwarzschild code may be able to find
models that fit the data outside our statistically allowed region. But
we also suspect that these models may be very unsmooth in phase space.
How a physical requirement of smoothness is best imposed on Schwarzschild
models is a matter of vigorous debate \citep{VME}.

The SAURON kinematic maps constitute the best data to date for revealing
the dynamical structure of elliptical galaxies. To fully realize the
potential of these data will require the application of a variety of
modeling approaches, including Schwarzschild's method, Jeans models, and
VF fitting. But it is not feasible to apply the most computationally expensive
method blindly; each approach should be informed by the results of coarser
methods that can survey more territory. This paper provides an initial
rough map of the parameter space for NGC~4365, as a guide for more detailed
modeling efforts to follow.

\section{Summary\label{S.SUMMARY}}

We have modeled the two-dimensional mean velocity field and isophotal profiles
of the old and kinematically interesting elliptical galaxy NGC~4365 using
the VF fitting approach, and constrained the system's intrinsic shape
between $0.03$ and $0.5$
effective radii and its orientation in space. We find that the galaxy is
highly triaxial ($\langle T \rangle \approx 0.45$) and somewhat flatter
than it appears ($\langle c/a \rangle \approx 0.6$). Axisymmetry or near
axisymmetry ($T<0.1$) is ruled out at better than 95\% confidence everywhere,
and at better than 99\% confidence everywhere except in the inner
$3\arcsec$. There is a weak indication of an outward triaxiality gradient. 
The orientation of the galaxy is constrained to two fairly narrow strips on
the viewing hemisphere. In the most probable models the system is oriented
with its long axis pointing in the general direction of the observer,
extending in projection to the southwest. We
have verified that the best models are statistically acceptable in a
$\chi^2$ sense, and that $\Delta\chi^2$ regions and Bayesian HPD regions
define essentially the same region of permitted parameter space.

That strong triaxiality has evidently persisted in the galaxy for some
$12\gyr$ implies that any central black hole must be smaller than $3
\times 10^9 \msun$. A larger central mass would have induced chaos
sufficient either to globally axisymmetrize the galaxy or at least make
the inner several arcseconds spherical. On the other hand, a black hole
of mass $4\times 10^8\msun$, as predicted by the $M_{\rm BH}$-$\sigma$
relation, would probably not preclude long-lived triaxiality and is
consistent with the observations.

We find that there must be a significant contribution from long-axis tubes
outside the kinematically decoupled core, and that the direct and retrograde
branches of these orbits must be unequally populated so as to produce
the observed rotation signature. This permits us to constrain the rate
of figure rotation about the short axis, since figure rotation
would induce an intrinsic twist. From the observed photometric twist we
infer that the tumbling period is $\gtrsim 500\myr$, putting corotation
at at least 8 effective radii, and supporting the common assumption that
figure rotation is dynamically unimportant in luminous ellipticals.

NGC~4365 contributes to an emerging picture of the role of black holes in
giant elliptical galaxies: though omnipresent, black holes are not massive
enough to alter the global structure of their hosts, which can remain
anisotropic and triaxial for many hundreds of dynamical times. It is
still undetermined whether this same picture may apply to lower-luminosity
systems, however, and it is possible that there is a threshold below which
central black holes drive a secular evolution toward axisymmetry. The
SAURON project has produced a rich database which will make it finally
possible to address these and related issues. Determining
the mass distributions and the dynamical structure of early-type galaxies
will require extensive modeling using Schwarzschild's method, as well as
faster but more idealized moment-based methods such as we have used.
We hope that through an application of such complementary approaches, it
will be possible to develop a more complete understanding of the dynamical
histories of hot stellar systems.

\section*{Acknowledgments}

This work was supported in part by NSF Career grant AST-9703036 to Ohio
University. Computations were performed using the facilities of the Ohio
Supercomputer Center and with facilities granted under the OSC Cluster Ohio
Program. TSS is grateful for the hospitality of the Observatoire de Lyon,
Sterrewacht Leiden, the School of Physics and Astronomy at the
University of Nottingham, and the Department of Astrophysics at the
University of Oxford, where parts of this work were carried out. The paper
was improved considerably by helpful advice from Tim de Zeeuw, Michele
Cappellari, Glenn van de Ven, and Tom Loredo.

\appendix

\section{Prior Distributions and Penalty Functions}

The separation of the prior distribution into factors for the orientation,
the parent mean shape distribution, shape correlations, and the
dynamical parameter distribution is explained in \S\ \ref{s.fitting}.
This appendix describes the details of the last two factors.

\subsection{Shape Prior}

The prior shape distribution for 8 radial bins is a joint distribution in
16 variables. It may seem least biased to let this function be constant
everywhere and allow any shape at each radius, but this is not the case.
The shape parameter space at each
radius is two-dimensional, and so a flat distribution implies that
large differences in shape between adjacent radii are likely
simply because there is more area in a plane far from a given point
than close to it. If the observations show that the galaxy appears
similar at neighboring radii---in particular, that the isophotal twist
is small---then the Bayesian ``robot'' \citep{Jaynes} is forced to reason
as follows: neighboring radii probably have very different shapes, yet
in projection they appear very similar; so we must be looking at the
galaxy along one of the special lines of sight where big changes of shape
are not noticeable. The results will then be biased toward
lines of sight in the symmetry planes.

From observations of many ellipticals it is already known that radial
photometric gradients tend to be small, and therefore that shape
profiles are probably smooth and shapes at neighboring radii are
correlated. Expected correlations must be included in the prior
\citep{Jaynes}. We do this by imposing a Gaussian penalty factor of the form
\beq
F_{\rm cor} = \prod_{i=2}^8 \exp \left\{- {(T_i-T_{i-1})^2
+ [(c/a)_i-(c/a)_{i-1}]^2 \over 2 \sigma_g^2 } \right\},
\eeq
which penalizes differences in $T$ and $c/a$ between neighboring radii
\citep[\S\ 3.2.2]{Statler3379}.
We set the constant $\sigma_g=0.126$, so as to render a complete
change of shape, e.g., from $T=0$ to $T=1$ or $c/a=0$ to $c/a=1$ over
the observed range of radii, a priori unlikely at the $3\sigma$ level.
Other values of $\sigma_g$ in the range $0.1$ to $0.2$ give very similar
results.

\subsection{Dynamical Prior}

The parent distribution of the dynamical parameters $C$, $k_X$, and
$k_Z$ is defined by the grid of values over which we compute models. For
the Z-tube/X-tube contrast we take
$C=(0,0.0625,0.125,0.25,0.5,1.0,2.0,\infty)$. The limiting values
$C=0$ and $C=\infty$ correspond to zero net streaming in Z-tubes
and X-tubes, respectively. We also add four more cases, where $C$ is
assumed to be a function of $T$ and $c/a$ according to each of the four
``prescription'' functions introduced in equations (11)--(14) of
\cite{Statler94c}. We find that the results from the regular grid alone
are nearly indistinguishable from those including the
prescriptions, indicating that explicit correlations between shape
and dynamics are unimportant.

The other constants $k_X$ and $k_Z$ determine the ``disc-like'' or
``spheroid-like'' character of the mean streaming in each tube family. A
lower value indicates more disc-like kinematics, meaning that the mean
velocity decreases more rapidly away from the symmetry plane. We take
$k_X=(0,0.25,0.5,0.75,1.0,1.25,1.5)$ and $k_Z=(0,0.25,0.5,0.75,1.0)$,
based on the behavior of numerical and analytic self-consistent triaxial
models [see Figure~8 of \cite{Statler94a}].

\subsection{$v/\sigma$ Penalty}

The magnitude of any velocity field can be scaled up or down by
reversing the direction of some fraction of the stars on tube orbits.
Accordingly, the model velocities are determined only to within a constant
factor and then scaled to fit the data. In previous work we have allowed
the scaling constant to vary freely. However, we find that this practice
leads to biased results when the apparent rotation axis is far from a
symmetry axis. The reason is that the fitting procedure can most easily
produce a model with the apparent rotation axis along an arbitrary
direction merely by placing the line of sight very close to the
true rotation axis, and then slightly inclining the model in the right
direction. But in this case the low $\sin i$ factor may imply
deprojected internal velocities that are unrealistically large.

To avoid these unphysical models, we calculate for each projected model
at each radius the maximum deprojected velocity on the shell, $v_s$, required
by the optimal scaling, and then penalize the model by a factor
$\exp(-v_s^2/2\sigma^2)$. We take the parameter $\sigma$ to be the
observed velocity dispersion at that radius, so in other words we are
penalizing models with internal $v/\sigma$ values greater than 1. This
turns out to be a fairly weak penalty in nearly all regions of parameter
space, affecting lines of sight only within about $15\arcdeg$ of the true
rotation axis.

\bsp
\label{lastpage}

\end{document}